\documentclass[runningheads]{llncs}

\usepackage{times}
\usepackage{graphicx}
\usepackage{acronym}
\usepackage{amssymb}
\usepackage{tabularx,amsmath}
\usepackage{multirow}
\usepackage{xcolor}
\usepackage{booktabs}
\usepackage[np, autolanguage]{numprint}
\usepackage{enumerate}
\usepackage{enumitem}
\usepackage[noend]{algorithmic}
\usepackage{algorithm}
\usepackage[caption=false]{subfig}

\newcommand{\dubbelop}{${\tiny \blacktriangle}$}

\newcommand{\dubbelneer}{${\tiny \blacktriangledown}$}

\acrodef{IR}{Information Retrieval}
\acrodef{LTR}{Learning to Rank}
\acrodef{OLTR}{Online Learning to Rank}
\acrodef{DBGD}{Dueling Bandit Gradient Descent}
\acrodef{MGD}{Multileave Gradient Descent}
\acrodef{C-MGD}{Cascading Multileave Gradient Descent}
\acrodef{RL}{Reinforcement Learning}
\acrodef{PL}{Plackett-Luce}

\acrodef{PUB-Rank}{Pairwise Unbiased Ranker Optimization}
\acrodef{PUGD}{Pairwise Unbiased Gradient Descent}
\acrodef{PDGD}{Pairwise Differentiable Gradient Descent}

\begin{document}

\title{On the Optimization of Ranking Models in the Online Setting}
\title{Optimizing Ranking Models in an Online Setting}

\author{Harrie Oosterhuis \and Maarten de Rijke}
\institute{
University of Amsterdam, Amsterdam, The Netherlands\\
\email{\{oosterhuis, derijke\}@uva.nl}}

\maketitle

\begin{abstract}
\ac{OLTR} methods optimize ranking models by directly interacting with users, which allows them to be very efficient and responsive.
All \ac{OLTR} methods introduced during the past decade have extended on the original \ac{OLTR} method: \ac{DBGD}.
Recently, a fundamentally different approach was introduced with the \ac{PDGD} algorithm.
To date the only comparisons of the two approaches are limited to simulations with cascading click models and low levels of noise.
The main outcome so far is that \ac{PDGD} converges at higher levels of performance and learns considerably faster than \ac{DBGD}-based methods.
However, the \ac{PDGD} algorithm assumes cascading user behavior, potentially giving it an unfair advantage.
Furthermore, the robustness of both methods to high levels of noise has not been investigated.
Therefore, it is unclear whether the reported advantages of \ac{PDGD} over \ac{DBGD} generalize to different experimental conditions.
In this paper, we investigate whether the previous conclusions about the \ac{PDGD} and \ac{DBGD} comparison generalize from ideal to worst-case circumstances.
We do so in two ways.
First, we compare the theoretical properties of \ac{PDGD} and \ac{DBGD}, by taking a critical look at previously proven properties in the context of ranking.
Second, we estimate an upper and lower bound on the performance of methods by simulating both \emph{ideal} user behavior and extremely \emph{difficult} behavior, i.e., almost-random non-cascading user models.
Our findings show that the theoretical bounds of \ac{DBGD} do not apply to any common ranking model and, furthermore, that the performance of \ac{DBGD} is substantially worse than \ac{PDGD} in both ideal and worst-case circumstances.
These results reproduce previously published findings about the relative performance of \ac{PDGD} vs.\ \ac{DBGD} and generalize them to extremely noisy and non-cascading circumstances.
\end{abstract}


\section{Introduction}
\label{sec:intro}
\ac{LTR} plays a vital role in information retrieval. It allows us to optimize models that combine hundreds of signals to produce rankings, thereby making large collections of documents accessible to users through effective search and recommendation.
Traditionally, \ac{LTR} has been approached as a supervised learning problem, where annotated datasets provide human judgements indicating relevance.
Over the years, many limitations of such datasets have become apparent: they are costly to produce~\cite{Chapelle2011,qin2013introducing} and actual users often disagree with the relevance annotations~\cite{sanderson2010}.
As an alternative, research into \ac{LTR} approaches that learn from user behavior has increased.
By learning from the implicit feedback in user behavior, users' true preferences can potentially be learned.
However, such methods must deal with the noise and biases that are abundant in user interactions~\cite{yue2010beyond}.
Roughly speaking, there are two approaches to \ac{LTR} from user interactions: learning from historical interactions and \acf{OLTR}\acused{OLTR}.
Learning from historical data allows for optimization without gathering new data~\cite{joachims2017unbiased}, though it does require good models of the biases in logged user interactions~\cite{chuklin-click-2015}.
In contrast, \ac{OLTR} methods learn by interacting with the user, thus they gather their own learning data.
As a result, these methods can adapt instantly and are potentially much more responsive than methods that use historical data.

\acfi{DBGD}~\cite{yue09:inter} \acused{DBGD} is the most prevalent \ac{OLTR} method; it has served as the basis of the field for the past decade.
\ac{DBGD} samples variants of its ranking model, and compares them using interleaving to find improvements~\cite{hofmann2011probabilistic,radlinski2013optimized}.
Subsequent work in \ac{OLTR} has extended on this approach~\cite{hofmann_2013_reusing,schuth2016mgd,wang2018efficient}.
Recently, the first alternative approach to \ac{DBGD} was introduced with \acfi{PDGD}\acused{PDGD}~\cite{Oosterhuis2018Unbiased}.
\ac{PDGD} estimates a pairwise gradient that is reweighed to be unbiased w.r.t.\ users' document pair preferences.
The original paper that introduced \ac{PDGD} showed considerable improvements over \ac{DBGD} under simulated user behavior~\cite{Oosterhuis2018Unbiased}: a substantially higher point of performance at convergence and a much faster learning speed.
The results in~\cite{Oosterhuis2018Unbiased} are based on simulations using low-noise cascading click models.
The pairwise assumption that \ac{PDGD} makes, namely, that all documents preceding a clicked document were observed by the user, is always correct in these circumstances, thus potentially giving it an unfair advantage over \ac{DBGD}.
Furthermore, the low level of noise presents a close-to-ideal situation, and it is unclear whether the findings in~\cite{Oosterhuis2018Unbiased} generalize to less perfect circumstances.

In this paper, we contrast \ac{PDGD} over \ac{DBGD}.
Prior to an experimental comparison, we determine whether there is a theoretical advantage of \ac{DBGD} over \ac{PDGD} and examine the regret bounds of \ac{DBGD} for ranking problems.
We then investigate whether the benefits of \ac{PDGD} over \ac{DBGD} reported in~\cite{Oosterhuis2018Unbiased} generalize to circumstances ranging from ideal to worst-case.
We simulate circumstances that are perfect for both methods -- behavior without noise or position-bias --  and circumstances that are the worst possible scenario -- almost-random, extremely-biased, non-cascading behavior.
These settings provide estimates of upper and lower bounds on performance, and indicate how well previous comparisons generalize to different circumstances.
Additionally, we introduce a version of \ac{DBGD} that is provided with an oracle interleaving method; its performance shows us the maximum performance \ac{DBGD} could reach from hypothetical extensions.

In summary, the following research questions are addressed in this paper:
\begin{enumerate}[align=left, label={\bf RQ\arabic*}, leftmargin=*, nosep, topsep=5pt]
    \item Do the regret bounds of \ac{DBGD} provide a benefit over \ac{PDGD}?\label{rq:regret}
    \item Do the advantages of \ac{PDGD} over \ac{DBGD} observed in prior work generalize to extreme levels of noise and bias? \label{rq:noisebias}
    \item Is the performance of \ac{PDGD} reproducible under non-cascading user behavior? \label{rq:cascading}
\end{enumerate}


\section{Related Work}
\label{sec:relatedwork}

This section provides a brief overview of traditional \ac{LTR} (Section~\ref{sec:ltr:annotated}), of \ac{LTR} from historical interactions (Section~\ref{sec:ltr:history}), and \ac{OLTR} (Section~\ref{sec:ltr:online}).

\subsection{Learning to rank from annotated datasets}
\label{sec:ltr:annotated}

Traditionally, \ac{LTR} has been approached as a supervised problem; in the context of \ac{OLTR} this approach is often referred to as \emph{offline} \ac{LTR}.
It requires a dataset containing relevance annotations of query-document pairs, after which a variety of methods can be applied \cite{liu2009:learning}.
The limitations of offline \ac{LTR} mainly come from obtaining such annotations.
The costs of gathering annotations are high as it is both time-consuming and expensive \cite{Chapelle2011,qin2013introducing}.
Furthermore, annotators cannot judge for very specific users, i.e., gathering data for personalization problems is infeasible.
Moreover, for certain applications it would be unethical to annotate items, e.g., for search in personal emails or documents \cite{wang2016learning}.
Additionally, annotations are stationary and cannot account for (perceived) relevance changes~\cite{dumais-web-2010,lefortier-online-2014,vakkari-changes-2000}.
Most importantly, though, annotations are not necessarily aligned with user preferences; judges often interpret queries differently from actual users~\cite{sanderson2010}.
As a result, there has been a shift of interest towards \ac{LTR} approaches that do not require annotated data.

\subsection{Learning to rank from historical interactions}
\label{sec:ltr:history}

The idea of \ac{LTR} from user interactions is long-established; one of the earliest examples is the original pairwise \ac{LTR} approach \cite{Joachims2002}.
This approach uses historical click-through interactions from a search engine and considers clicks as indications of relevance.
Though very influential and quite effective, this approach ignores the \emph{noise} and \emph{biases} inherent in user interactions.
Noise, i.e., any user interaction that does not reflect the user's true preference, occurs frequently, since many clicks happen for unexpected reasons~\cite{sanderson2010}.
Biases are systematic forms of noise that occur due to factors other than relevance.
For instance, interactions will only involve displayed documents resulting in selection bias~\cite{wang2016learning}.
Another important form of bias in \ac{LTR} is position bias, which occurs because users are less likely to consider documents that are ranked lower~\cite{yue2010beyond}.
Thus, to learn true preferences from user interactions effectively, a \ac{LTR} method should be robust to noise and handle biases correctly.

In recent years counter-factual \ac{LTR} methods have been introduced that correct for some of the bias in user interactions.
Such methods uses inverse propensity scoring to account for the probability that a user observed a ranking position~\cite{joachims2017unbiased}.
Thus, clicks on positions that are observed less often due to position bias will have greater weight to account for that difference.
However, the position bias must be learned and estimated somewhat accurately~\cite{ai2018unbiased}.
On the other side of the spectrum are click models, which attempt to model user behavior completely~\cite{chuklin-click-2015}.
By predicting behavior accurately, the effect of relevance on user behavior can also be estimated~\cite{borisov2016neural,wang2016learning}.

An advantage of these approaches over \ac{OLTR} is that they only require historical data and thus no new data has to be gathered. However, unlike \ac{OLTR}, they do require a fairly accurate user model, and thus they cannot be applied in cold-start situations.

\subsection{Online learning to rank}
\label{sec:ltr:online}

\ac{OLTR} differs from the approaches listed above because its methods intervene in the search experience.
They have control over what results are displayed, and can learn from their interactions instantly.
Thus, the online approach performs \ac{LTR} by interacting with users directly~\cite{yue09:inter}.
Similar to \ac{LTR} methods that learn from historical interaction data, \ac{OLTR} methods have the potential to learn the true user preferences.
However, they also have to deal with the noise and biases that come with user interactions.
Another advantage of \ac{OLTR} is that the methods are very responsive, as they can apply their learned behavior instantly.
Conversely, this also brings a danger as an online method that learns incorrect preferences can also worsen the experience immediately.
Thus, it is important that \ac{OLTR} methods are able to learn reliably in spite of noise and biases.
Thus, \ac{OLTR} methods have a two-fold task: they have to simultaneously present rankings that provide a good user experience \emph{and} learn from user interactions with the presented rankings.

The original \ac{OLTR} method is \acf{DBGD}; it approaches optimization as a dueling bandit problem~\cite{yue09:inter}.
This approach requires an online comparison method that can compare two rankers w.r.t. user preferences; traditionally, \ac{DBGD} methods use interleaving.
Interleaving methods take the rankings produced by two rankers and combine them in a single result list, which is then displayed to users.
From a large number of clicks on the presented list the interleaving methods can reliably infer a preference between the two rankers~\cite{hofmann2011probabilistic,radlinski2013optimized}.
At each timestep, \ac{DBGD} samples a candidate model, i.e., a slight variation of its current model, and compares the current and candidate models using interleaving.
If a preference for the candidate is inferred, the current model is updated towards the candidate slightly.
By doing so, \ac{DBGD} will update its model continuously and should oscillate towards an inferred optimum.
Section~\ref{sec:DBGD} provides a complete description of the \ac{DBGD} algorithm.

Virtually all work in \ac{OLTR} in the decade since the introduction of \ac{DBGD} has used \ac{DBGD} as a basis.
A straightforward extension comes in the form of Multileave Gradient Descent~\cite{schuth2016mgd} which compares a large number of candidates per interaction~\cite{oosterhuis2017sensitive,schuth2015probabilistic,Schuth2014a}.
This leads to a much faster learning process, though in the long term this method does not seem to improve the point of convergence.

One of the earliest extensions of \ac{DBGD} proposed a method for reusing historical interactions to guide exploration for faster learning \cite{hofmann_2013_reusing}.
While the initial results showed great improvements~\cite{hofmann_2013_reusing}, later work showed performance drastically decreasing in the long term due to bias introduced by the historical data~\cite{oosterhuis2016probabilistic}.
Unfortunately, \ac{OLTR} work that continued this historical approach~\cite{wang2018efficient} also only considered short term results; moreover, the results of some work~\cite{zhao2016constructing} are not based on held-out data.
As a result, we do not know whether these extensions provide decent long-term performance and it is unclear whether the findings of these studies generalize to more realistic settings.

Recently, an inherently different approach to \ac{OLTR} was introduced with \ac{PDGD}~\cite{Oosterhuis2018Unbiased}.
\ac{PDGD} interprets its ranking model as a distribution over documents; it estimates a pairwise gradient from user interactions with sampled rankings.
This gradient is differentiable, allowing for non-linear models like neural networks to be optimized, something \ac{DBGD} is ineffective at~\cite{oosterhuis2017balancing,Oosterhuis2018Unbiased}.
Section~\ref{sec:PDGD} provides a detailed description of \ac{PDGD}.
In the paper in which we introduced \ac{PDGD}, claim that it provides substantial improvements over \ac{DBGD}.
However, those claims are based on cascading click models with low levels of noise.
This is problematic because \ac{PDGD} assumes a cascading user, and could thus have an unfair advantage in this setting.
Furthermore, it is unclear whether \ac{DBGD} with a perfect interleaving method could still improve over \ac{PDGD}.
Lastly, \ac{DBGD} has proven regret bounds while \ac{PDGD} has no such guarantees.

In this study, we clear up these questions about the relative strengths of \ac{DBGD} and \ac{PDGD} by comparing the two methods under non-cascading, high-noise click models.
Additionally, by providing \ac{DBGD} with an oracle comparison method, its hypothetical maximum performance can be measured; thus, we can study whether an improvement over \ac{PDGD} is hypothetically possible.
Finally, a brief analysis of the theoretical regret bounds of \ac{DBGD} shows that they do not apply to any common ranking model, therefore hardly providing a guaranteed advantage over \ac{PDGD}.


\section{Dueling Bandit Gradient Descent}
\label{sec:DBGD}

This section describes the \ac{DBGD} algorithm in detail, before discussing the regret bounds of the algorithm.

\begin{algorithm}[t]
\caption{\acf{DBGD}.} 
\label{alg:dbgd}
\begin{algorithmic}[1]
\STATE \textbf{Input}: initial weights: $\theta_1$; unit: $u$; learning rate $\eta$.  \label{line:dbgd:initmodel}
\FOR{$t \leftarrow  1 \ldots \infty$ }
	\STATE $q_t \leftarrow \mathit{receive\_query}(t)$\hfill \textit{\small  obtain a query from a user} \label{line:dbgd:query}
	\STATE $\theta_t^{c} \gets \theta_t +  \mathit{sample\_from\_unit\_sphere}(u)$  \hfill \textit{\small  create candidate ranker} \label{line:dbgd:candidate}
	\STATE $R_t \leftarrow \mathit{get\_ranking(\theta_t, D_{q_t})}$  \hfill \textit{\small  get current ranker ranking} \label{line:dbgd:ranking}
	\STATE $R_t^c \leftarrow \mathit{get\_ranking(\theta_t^c, D_{q_t})}$  \hfill \textit{\small  get candidate ranker ranking} \label{line:dbgd:candidateranking}
	\STATE $I_t \leftarrow \mathit{interleave(R_t, R_t^c)}$ \hfill \textit{\small  interleave both rankings} \label{line:dbgd:interleave}
	\STATE $\mathbf{c}_t \leftarrow \mathit{display\_to\_user}(I_t)$ \hfill \textit{\small  displayed interleaved list, record clicks} \label{line:dbgd:display}
	\IF{$\mathit{preference\_for\_candidate}(I_t, \mathbf{c}_t, R_t, R_t^c)$}
    		\STATE $\theta_{t+1} \leftarrow \theta_t + \eta (\theta^c_t - \theta_t)$ \hfill \textit{\small  update model towards candidate} \label{line:dbgd:update}
	\ELSE
		\STATE $\theta_{t+1} \leftarrow \theta_t$ \hfill \textit{\small  no update} \label{line:dbgd:noupdate}
	\ENDIF
\ENDFOR
\end{algorithmic}
\end{algorithm}

\subsection{The \acl{DBGD} method}

The \ac{DBGD} algorithm~\cite{yue09:inter} describes an indefinite loop that aims to improve a ranking model at each step; Algorithm~\ref{alg:dbgd} provides a formal description.
The algorithm starts a given model with weights $\theta_1$ (Line~\ref{line:dbgd:initmodel}); then it waits for a user-submitted query (Line~\ref{line:dbgd:query}).
At this point a candidate ranker is sampled from the unit sphere around the current model (Line~\ref{line:dbgd:candidate}), and the current and candidate model both produce a ranking for the current query (Line~\ref{line:dbgd:ranking}~and~\ref{line:dbgd:candidateranking}).
These rankings are interleaved (Line~\ref{line:dbgd:interleave}) and displayed to the user (Line~\ref{line:dbgd:display}).
If the interleaving method infers a preference for the candidate ranker from subsequent user interactions the current model is updated towards the candidate (Line~\ref{line:dbgd:update}), otherwise no update is performed (Line~\ref{line:dbgd:noupdate}).
Thus, the model optimized by \ac{DBGD} should converge and oscillate towards an optimum.

\subsection{Regret bounds of \acl{DBGD}}
\label{subsection:regretbounds}
 
Unlike \ac{PDGD}, \ac{DBGD} has proven regret bounds~\cite{yue09:inter}, potentially providing an advantage in the form of theoretical guarantees.
In this section we answer \ref{rq:regret} by critically looking at the assumptions which form the basis of \ac{DBGD}'s proven regret bounds.

The original \ac{DBGD} paper~\cite{yue09:inter} proved a sublinear regret under several assumptions.
\ac{DBGD} works with the parameterized space of ranking functions $\mathcal{W}$, that is, every $\theta \in \mathcal{W}$ is a different set of parameters for a ranking function.
For this study we will only consider linear models because all existing \ac{OLTR} work has dealt with them \cite{hofmann_2013_reusing,hofmann11:balancing,Oosterhuis2018Unbiased,oosterhuis2016probabilistic,schuth2016mgd,wang2018efficient,yue09:inter,zhao2016constructing}.
But we note that the proof is easily extendable to neural networks where the output is a monotonic function applied to a linear combination of the last layer.
Then there is assumed to be a concave utility function $u : \mathcal{W} \rightarrow \mathbb{R}$; since this function is concave, there should only be a single instance of weights that are optimal $\theta^*$.
Furthermore, this utility function is assumed to be L-Lipschitz smooth:
\begin{align}
\exists L \in \mathbb{R},\quad \forall (\theta_a, \theta_b) \in \mathcal{W},\quad |u(\theta_a) - u(\theta_b)| < L \|\theta_a - \theta_b \|.
\end{align}
We will show that these assumptions are \emph{incorrect}: there is an infinite number of optimal weights, and the utility function $u$ cannot be L-Lipschitz smooth.
Our proof relies on two assumptions that avoid cases where the ranking problem is trivial.
First, the zero ranker is not the optimal model:
\begin{align}
\theta^* \not= \mathbf{0}.
\end{align}
Second, there should be at least two models with different utility values:
\begin{align}
\exists (\theta, \theta') \in \mathcal{W},\quad u(\theta) \not= u(\theta').
\end{align}
We will start by defining the set of rankings a model $f(\cdot, \theta)$ will produce as:
\begin{align}
\mathcal{R}_D(f(\cdot, \theta))
=
\{ R \mid \forall (d, d') \in D, [f(d, \theta) > f(d', \theta) \rightarrow d \succ_R d']\} .
\end{align}
It is easy to see that multiplying a model with a positive scalar $\alpha > 0$ will not affect this set:
\begin{align}
\forall \alpha \in \mathbb{R}_{>0},\quad  \mathcal{R}_D(f(\cdot, \theta))
= \mathcal{R}_D(\alpha f(\cdot, \theta)).
\end{align}
Consequently, the utility of both functions will be equal:
\begin{align}
\forall \alpha \in \mathbb{R}_{>0},\quad  u(f(\cdot, \theta))
= u(\alpha f(\cdot, \theta)).
\end{align}
For linear models scaling weights has the same effect: $\alpha f(\cdot, \theta) = f(\cdot, \alpha \theta)$.
Thus, the first assumption cannot be true since for any optimal model $f(\cdot, \theta^*)$ there is an infinite set of equally optimal models: $\{f(\cdot, \alpha \theta^*) \mid \alpha \in \mathbb{R}_{>0}\}$.

Then, regarding L-Lipschitz smoothness, using any positive scaling factor:
\begin{align}
\forall \alpha \in \mathbb{R}_{>0},\quad& |u(\theta_a) - u(\theta_b)| = |u(\alpha\theta_a) - u(\alpha\theta_b)|, \\
\forall \alpha \in \mathbb{R}_{>0},\quad& \| \alpha\theta_a - \alpha\theta_b \| = \alpha \| \theta_a - \theta_b \|.
\end{align}
Thus the smoothness assumption can be rewritten as:
\begin{align}
\exists L \in \mathbb{R},\quad \forall \alpha \in \mathbb{R}_{>0},\quad \forall (\theta_a, \theta_b) \in \mathcal{W},\quad |u(\theta_a) - u(\theta_b)| < \alpha L\| \theta_a - \theta_b \|.
\end{align}
However, there is always an infinite number of values for $\alpha$ small enough to break the assumption.
Therefore, we conclude that a concave L-Lipschitz smooth utility function can never exist for a linear ranking model, thus the proof for the regret bounds is not applicable when using linear models.

Consequently, the regret bounds of \ac{DBGD} do not apply to the ranking problems in previous work.
One may consider other models (e.g., spherical coordinate based models), however this still means that for the simplest and most common ranking problems there are no proven regret bounds.
As a result, we answer \ref{rq:regret} negatively, the regret bounds of \ac{DBGD} do not provide a benefit over \ac{PDGD} for the ranking problems in \ac{LTR}.


\section{Pairwise Differentiable Gradient Descent}
\label{sec:PDGD}

The \acf{PDGD}~\cite{Oosterhuis2018Unbiased} algorithm is formally described in Algorithm~\ref{alg:pdgd}.
\ac{PDGD} interprets a ranking function $f(\cdot, \theta)$ as a probability distribution over documents by applying a Plackett-Luce model:
\begin{align}
P(d | D, \theta) = \frac{e^{f(d,\theta)}}{\sum_{d' \in D} e^{f(d',\theta)}}. \label{eq:docprob}
\end{align}
First, the algorithm waits for a user query (Line~\ref{line:pdgd:query}), then a ranking $R$ is created by sampling documents without replacement (Line~\ref{line:pdgd:samplelist}).
Then \ac{PDGD} observes clicks from the user and infers pairwise document preferences from them.
All documents preceding  a clicked document and the first succeeding one are assumed to be observed by the user.
Preferences between clicked and unclicked observed documents are inferred by \ac{PDGD}; this is a long-standing assumption in pairwise \ac{LTR} \cite{Joachims2002}.
We denote an \emph{inferred} preference between documents as $d_i \succ_{\mathbf{c}} d_j$, and the probability of the model placing $d_i$ earlier than $d_j$ is denoted and calculated by:
\begin{align}
P(d_i \succ d_j \mid \theta) = \frac{e^{f(d_i,\theta)}}{e^{f(d_i,\theta)} + e^{f(d_j,\theta)}}.
\end{align}
The gradient is estimated as a sum over inferred preferences with a weight $\rho$ per pair:
\begin{equation}
\begin{split}
\mbox{}\hspace*{-0.6cm}\Delta &f(\cdot, \theta) \\
&\approx \sum_{d_i \succ_{\mathbf{c}} d_j} \rho(d_i, d_j, R, D) [\Delta P(d_i \succ d_j \mid \theta)]  \\[1.1ex]
&= \sum_{d_i \succ_{\mathbf{c}} d_j} \rho(d_i, d_j, R, D) P(d_i \succ d_j \mid \theta)P(d_j \succ d_i \mid \theta)(f'({d}_i, \theta) - f'({d}_j, \theta)).\hspace*{-1cm}\mbox{}
\end{split}
\label{eq:novelgradient}
\end{equation}
After computing the gradient (Line~\ref{line:pdgd:modelgrad}), the model is updated accordingly (Line \ref{line:pdgd:update}).
This will change the distribution (Equation~\ref{eq:docprob}) towards the inferred preferences.
This distribution models the confidence over which documents should be placed first; the exploration of \ac{PDGD} is naturally guided by this confidence and can vary per query.

\begin{algorithm}[t]
\caption{\acf{PDGD}.} 
\label{alg:pdgd}
\begin{algorithmic}[1]
\STATE \textbf{Input}: initial weights: $\mathbf{\theta}_1$; scoring function: $f$; learning rate $\eta$.  \label{line:pdgd:initmodel}
\FOR{$t \leftarrow  1 \ldots \infty$ }
	\STATE $q_t \leftarrow \mathit{receive\_query}(t)$\hfill \textit{\small // obtain a query from a user} \label{line:pdgd:query}
	\STATE $\mathbf{R}_t \leftarrow \mathit{sample\_list}(f_{\theta_t}, D_{q_t})$ \hfill \textit{\small // sample list according to Eq.~\ref{eq:docprob}} \label{line:pdgd:samplelist}
	\STATE $\mathbf{c}_t \leftarrow \mathit{receive\_clicks}(\mathbf{R}_t)$ \hfill \textit{\small // show result list to the user} \label{line:pdgd:clicks}
	\STATE $\nabla  f(\cdot,\theta_t) \leftarrow \mathbf{0}$ \hfill \textit{\small // initialize gradient} \label{line:pdgd:initgrad}
	\FOR{$d_i \succ_{\mathbf{c}} d_j \in \mathbf{c}_t$} \label{line:pdgd:prefinfer}
	\STATE $w \leftarrow \rho(d_i, d_j, R, D)$  \hfill \textit{\small // initialize pair weight (Eq.~\ref{eq:rho})} \label{line:pdgd:initpair}
	\STATE $w \leftarrow w \times  P(d_i \succ d_j \mid \theta_t)P(d_j \succ d_i \mid \theta_t)$
             \hfill \textit{\small // pair gradient (Eq.~\ref{eq:novelgradient})} \label{line:pdgd:pairgrad}
	\STATE  $\nabla  f(\cdot,\theta_t) \leftarrow \nabla f_{\theta_t} + w \times (f'({d}_i, \theta_t) - f'({d}_j, \theta_t))$
	  \hfill \textit{\small // model gradient (Eq.~\ref{eq:novelgradient})} \label{line:pdgd:modelgrad}
	\ENDFOR
	\STATE $\theta_{t+1} \leftarrow \theta_{t} + \eta \nabla  f(\cdot,\theta_t)$
	\hfill \textit{\small // update the ranking model} \label{line:pdgd:update}
\ENDFOR
\end{algorithmic}
\end{algorithm}

The weighting function $\rho$ is used to make the gradient of \ac{PDGD} unbiased w.r.t. document pair preferences.
It uses the reverse pair ranking: $R^*(d_i, d_j, R)$, which is the same ranking as $R$ but with the document positions of $d_i$ and $d_j$ swapped.
Then $\rho$ is the ratio between the probability of $R$ and $R^*$:
\begin{align}
\rho(d_i, d_j, R, D) &= \frac{P(R^*(d_i, d_j, R) \mid D)}{P(R \mid D) + P(R^*(d_i, d_j, R) \mid D)}. \label{eq:rho}
\end{align}
In the original \ac{PDGD} paper~\cite{Oosterhuis2018Unbiased}, the weighted gradient is proven to be unbiased w.r.t. document pair preferences under certain assumptions about the user.
Here, this unbiasedness is defined by being able to rewrite the gradient as:
\begin{align}
E[\Delta f(\cdot, \theta)] = \sum_{(d_i, d_j) \in D} \alpha_{ij}(f'(\mathbf{d}_i, \theta) - f'(\mathbf{d}_j, \theta)), \label{eq:unbias}
\end{align}
and the sign of $\alpha_{ij}$ agreeing with the preference of the user:
\begin{align}
\mathit{sign}(\alpha_{ij}) = \mathit{sign}(\mathit{relevance}(d_i) - \mathit{relevance}(d_j)). \label{eq:signunbias}
\end{align}
The proof in \cite{Oosterhuis2018Unbiased} only relies on the difference in the probabilities of inferring a preference: $d_i \succ_{\mathbf{c}} d_j$ in $R$ and the opposite preference $d_j \succ_{\mathbf{c}} d_i$ in $R^*(d_i, d_j, R)$. 
The proof relies on the sign of this difference to match the user's preference:
\begin{align}
\begin{split}
\mathit{sign}(P(d_i \succ_{\mathbf{c}} d_j \mid R) & - P(d_j \succ_{\mathbf{c}} d_i \mid R^*)) = {}\\
 &\mathit{sign}(\mathit{relevance}(d_i) - \mathit{relevance}(d_j)). \label{eq:signclick}
 \end{split}
\end{align}
As long as Equation~\ref{eq:signclick} is true, Equation~\ref{eq:unbias}~and~\ref{eq:signunbias} hold as well.
Interestingly, this means that other assumptions about the user can be made than in \cite{Oosterhuis2018Unbiased}, and other variations of \ac{PDGD} are possible, e.g., the algorithm could assume that all documents are observed and the proof still holds.

The original paper on \ac{PDGD} reports large improvements over \ac{DBGD}, however these improvements were observed under simulated cascading user models.
This means that the assumption that \ac{PDGD} makes about which documents are observed are always true.
As a result, it is currently unclear whether the method is really better in cases where the assumption does not hold.


\begin{table}[tb]
\caption{Click probabilities for simulated \emph{perfect} or \emph{almost random} behavior.}
\centering
\begin{tabularx}{\columnwidth}{ l X X X X X  }
\toprule
& \multicolumn{5}{c}{ $P(\mathit{click}(d)\mid \mathit{relevance}(d), \mathit{observed}(d))$} \\
\cmidrule(lr){2-6} 
$\mathit{relevance}(d)$ & \emph{$ 0$} & \emph{$ 1$}  &  \emph{$ 2$} & \emph{$ 3$} & \emph{$ 4$} \\
\midrule
 \emph{perfect}                           &  0.00 &  0.20 &  0.40 &  0.80 &  1.00  \\
  \emph{almost random}  \quad\quad\quad\quad &  0.40 &  0.45 &  0.50 &  0.55 &  0.60  \\
\bottomrule
\end{tabularx}
\label{tab:clickmodels}
\end{table}

\section{Experiments}
\label{sec:experiments}

In this section we detail the experiments that were performed to answer the research questions in Section~\ref{sec:intro}.\footnote{The resources for reproducing the experiments in this paper are available at \url{https://github.com/HarrieO/OnlineLearningToRank}}

\subsection{Datasets}
\label{sec:experiments:datasets}

Our experiments are performed over three large labelled datasets from commercial search engines, the largest publicly available \ac{LTR} datasets.
These datasets are the \emph{MLSR-WEB10K}~\cite{qin2013introducing}, \emph{Yahoo!\ Webscope}~\cite{Chapelle2011}, and \emph{Istella}~\cite{dato2016fast} datasets.
Each contains a set of queries with corresponding preselected document sets.
Query-document pairs are represented by feature vectors and five-grade relevance annotations ranging from \emph{not relevant} (0) to \emph{perfectly relevant}~(4).
Together, the datasets contain over \numprint{29900} queries and between 136 and 700~features per representation.

\subsection{Simulating user behavior}
\label{sec:experiments:users}

In order to simulate user behavior we partly follow the standard setup for \ac{OLTR}~\cite{He2009,hofmann11:balancing,oosterhuis2016probabilistic,schuth2016mgd,zoghi:wsdm14:relative}.
At each step a user issued query is simulated by uniformly sampling from the datasets.
The algorithm then decides what result list to display to the user, the result list is limited to $k = 10$ documents.
Then user interactions are simulated using click models~\cite{chuklin-click-2015}.
Past \ac{OLTR} work has only considered \emph{cascading click models}~\cite{guo09:efficient}; in contrast, we also use \emph{non-cascading click models}.
The probability of a click is conditioned on relevance and observance:
\begin{align}
P(\mathit{click}(d)\mid \mathit{relevance}(d), \mathit{observed}(d)).
\end{align}
We use two levels of noise to simulate \emph{perfect} user behavior and \emph{almost random} behavior~\cite{Hofmann2013a}, Table~\ref{tab:clickmodels} lists the probabilities of both.
The \emph{perfect} user observes all documents, never clicks on anything non-relevant, and always clicks on the most relevant documents.
Two variants of \emph{almost random} behavior are used.
The first is based on cascading behavior, here the user first observes the top document, then decides to click according to Table~\ref{tab:clickmodels}.
If a click occurs, then, with probability $P(stop \mid click) = 0.5$ the user stops looking at more documents, otherwise the process continues on the next document.
The second \emph{almost random} behavior is simulated in a non-cascading way; here we follow~\cite{joachims2017unbiased} and model the observing probabilities as:
\begin{align}
P(\mathit{observed}(d) \mid \mathit{rank}(d)) = \frac{1}{rank(d)}.
\end{align}
The important distinction is that it is safe to assume that the cascading user has observed all documents ranked before a click, while this is not necessarily true for the non-cascading user.
Since \ac{PDGD} makes this assumption, testing under both models can show us how much of its performance relies on this assumption.
Furthermore, the \emph{almost random} model has an extreme level of noise and position bias compared to the click models used in previous \ac{OLTR} work~\cite{hofmann11:balancing,oosterhuis2016probabilistic,schuth2016mgd}, and we argue it simulates an (almost) worst-case scenario.

\subsection{Experimental runs}
\label{sec:experiments:runs}

In our experiments we simulate runs consisting of \numprint{1000000} impressions; each run was repeated 125 times under each of the three click models.
\ac{PDGD} was run with $\eta = 0.1$ and zero initialization, \ac{DBGD} was run using Probabilistic Interleaving~\cite{oosterhuis2016probabilistic} with zero initialization, $\eta = 0.001$, and the unit sphere with $\delta=1$.
Other variants like Multileave Gradient Descent~\cite{schuth2016mgd} are not included; previous work has shown that their performance matches that of regular \ac{DBGD} after around \numprint{30000} impressions~\cite{Oosterhuis2018Unbiased,oosterhuis2016probabilistic,schuth2016mgd}.
The initial boost in performance comes at a large computational cost, though, as the fastest approaches keep track of at least 50 ranking models~\cite{oosterhuis2016probabilistic}, which makes running long experiments extremely impractical.
Instead, we introduce a novel oracle version of \ac{DBGD}, where, instead of interleaving, the NDCG values on the current query are calculated and the highest scoring model is selected.
This simulates a hypothetical perfect interleaving method, and we argue that the performance of this oracle run indicates what the upper bound on \ac{DBGD} performance is.

Performance is measured by NDCG@10 on a held-out test set, a two-sided t-test is performed for significance testing.
We do not consider the user experience during training, because past work has already investigated this aspect thoroughly~\cite{Oosterhuis2018Unbiased}.


\section{Experimental Results and Analysis}
\label{sec:results}

Recall that in Section~\ref{subsection:regretbounds} we have already provided a negative answer to \ref{rq:regret}: the regret bounds of \ac{DBGD} do not provide a benefit over \ac{PDGD} for the ranking problems in \ac{LTR}.
In this section we present our experimental results and answer \ref{rq:noisebias} (whether the advantages of \ac{PDGD} over \ac{DBGD} of previous work generalize to extreme levels of noise and bias) and \ref{rq:cascading} (whether the performance of \ac{PDGD} is reproducible under non-cascading user behavior). 

\begin{figure}[t]
\caption{Performance (NDCG@10) on held-out data from Yahoo (top), MSLR (center), Istella (bottom) datasets, under the \emph{perfect}, and \emph{almost random} user models: cascading (casc.) and non-cascading (non-casc.).
The shaded areas display the standard deviation.}
\begin{tabular}{l@{}l}
\includegraphics[width=.8\textwidth]{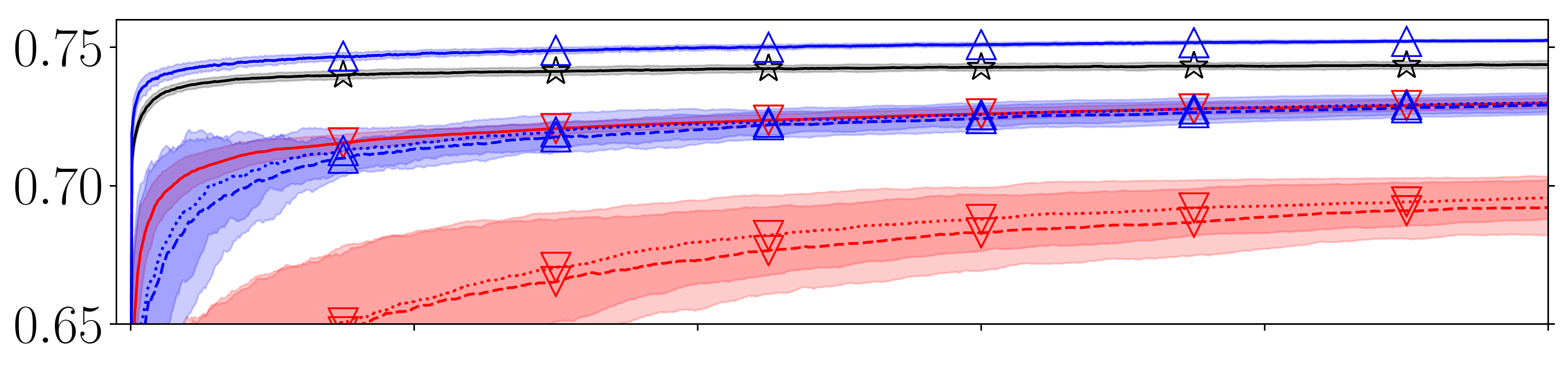} & \multirow{3}{*}[6em]{\includegraphics[width=.2\textwidth]{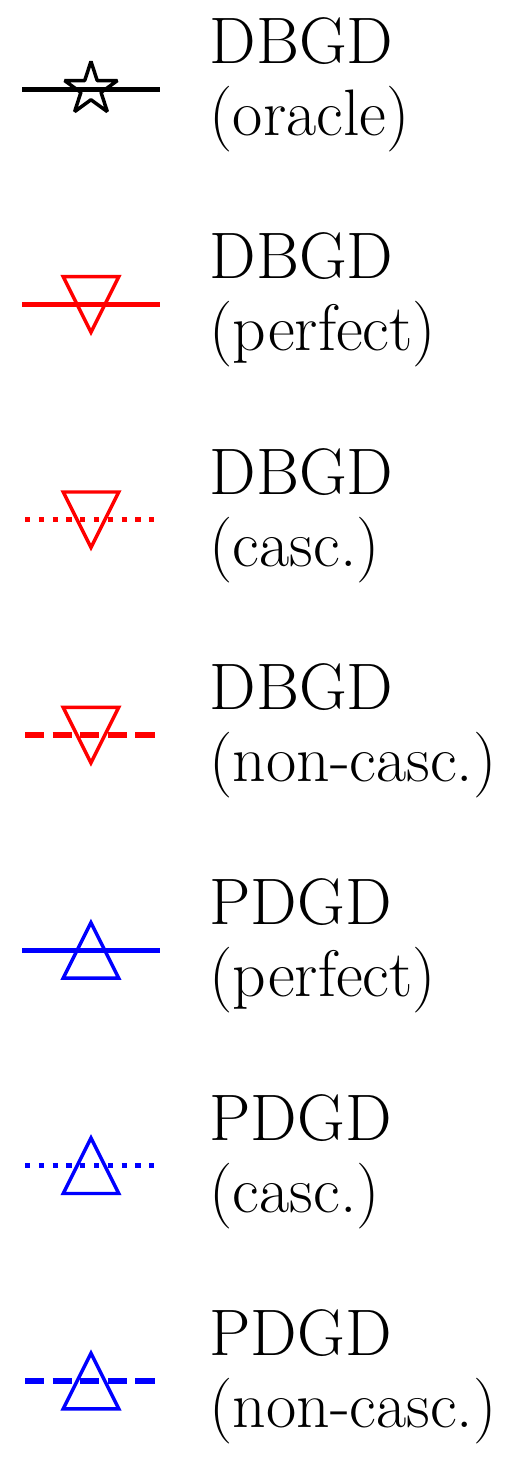}} \\
\hspace*{0.142em}
\includegraphics[width=.79\textwidth]{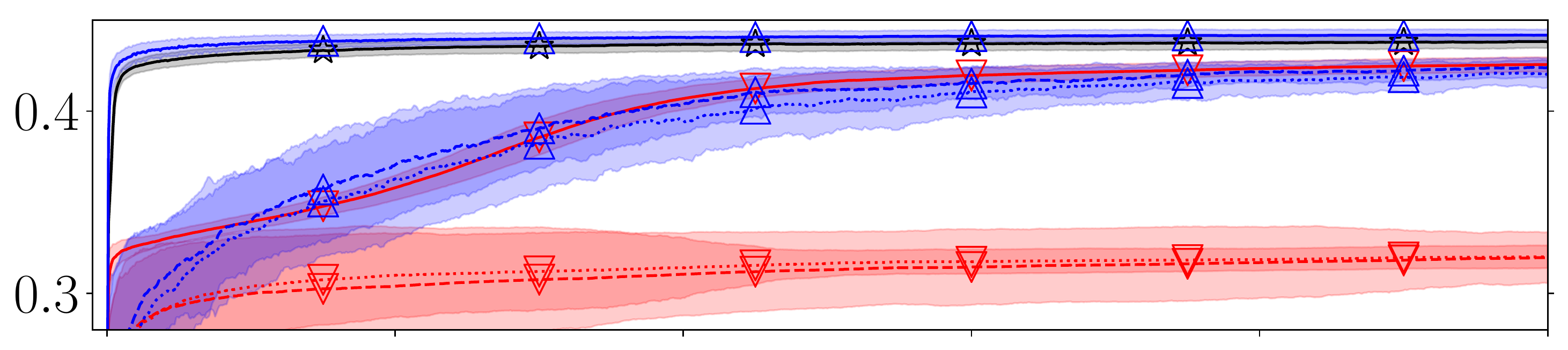} & \\
\hspace*{0.47em}%
\includegraphics[width=.8\textwidth]{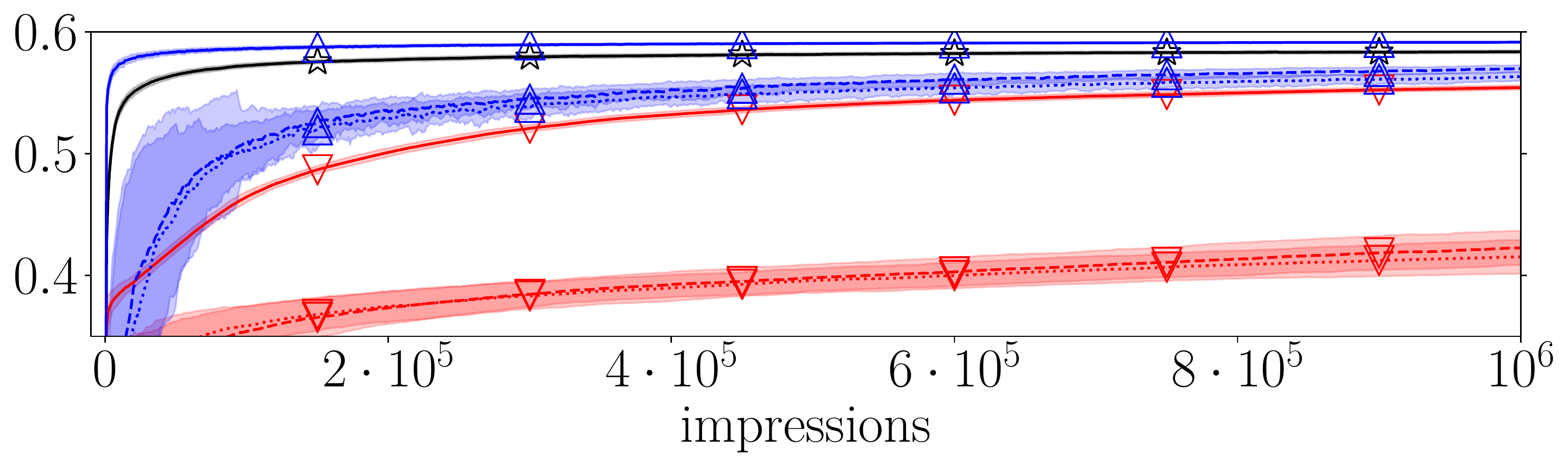} &
\end{tabular}
\label{fig:main}
\end{figure}
Our main results are presented in Table~\ref{tab:main}. Additionally, Figure~\ref{fig:main} displays the average performance over \numprint{1000000} impressions.
First, we consider the performance of \ac{DBGD}; there is a substantial difference between its performance under the \emph{perfect} and \emph{almost random} user models on all datasets.
Thus, it seems that \ac{DBGD} is strongly affected by noise and bias in interactions; interestingly, there is little difference between performance under the cascading and non-cascading behavior.
On all datasets the \emph{oracle} version of \ac{DBGD} performs significantly better than \ac{DBGD} under \emph{perfect} user behavior.
This means there is still room for improvement and hypothetical improvements in, e.g., interleaving could lead to significant increases in long-term \ac{DBGD} performance.

Next, we look at the performance of \ac{PDGD}; here, there is also a significant difference between performance under the \emph{perfect} and \emph{almost random} user models on all datasets.
However, the effect of noise and bias is very limited compared to \ac{DBGD}, and this difference at \numprint{1000000} impressions is always less than $0.03$ NDCG on any dataset.

To answer \ref{rq:noisebias}, we compare the performance of \ac{DBGD} and \ac{PDGD}.
Across all datasets, when comparing \ac{DBGD} and \ac{PDGD} under the same levels of interaction noise and bias, the performance of \ac{PDGD} is significantly better in every case.
Furthermore, \ac{PDGD} under the \emph{perfect} user model significantly outperforms the \emph{oracle} run of \ac{DBGD}, despite the latter being able to directly observe the NDCG of rankers on the current query.
Moreover, when comparing \ac{PDGD} performance under the \emph{almost random} user model with \ac{DBGD} under the \emph{perfect} user model, we see the differences are limited and in both directions.
Thus, even under ideal circumstances \ac{DBGD} does not consistently outperform \ac{PDGD} under extremely difficult circumstances.
As a result, we answer \ref{rq:noisebias} positively: our results strongly indicate that the performance of \ac{PDGD} is considerably better than \ac{DBGD} and that these findings generalize from ideal circumstances to settings with extreme levels of noise and bias.

Finally, to answer \ref{rq:cascading}, we look at the performance under the two \emph{almost random} user models.
Surprisingly, there is no clear difference between the performance of \ac{PDGD} under \emph{cascading} and \emph{non-cascading} user behavior.
The differences are small and per dataset it differs which circumstances are slightly preferred.
Therefore, we answer \ref{rq:cascading} positively: the performance of \ac{PDGD} is reproducible under \emph{non-cascading} user behavior.

\begin{table*}[t]
\centering
\caption{Performance (NDCG@10) after \numprint{1000000} impressions for \ac{DBGD} and \ac{PDGD} under a \emph{perfect} click model and two almost-random click models: \emph{cascading} and \emph{non-cascading}, and \ac{DBGD} with an \emph{oracle} comparator.
Significant improvements and losses (p $<$ 0.01) between \ac{DBGD} and \ac{PDGD} are indicated by \dubbelop, \dubbelneer, and $\circ$ (no significant difference). Indications are in order of: \emph{oracle}, \emph{perfect}, \emph{cascading}, and \emph{non-cascading}.
}
\begin{tabular*}{\textwidth}{@{\extracolsep{\fill} } l  l l l  }
\toprule
 & { \small \textbf{Yahoo}}  & { \small \textbf{MSLR}}  & { \small \textbf{Istella}} \\
\midrule
& \multicolumn{3}{c}{\textit{\acl{DBGD}}} \\
\midrule
\textit{oracle} & 0.744 {\tiny (0.001)} {\tiny \dubbelneer} {\tiny \dubbelop} {\tiny \dubbelop} & 0.438 {\tiny (0.004)} {\tiny \dubbelneer} {\tiny \dubbelop} {\tiny \dubbelop} & 0.584 {\tiny (0.001)} {\tiny \dubbelneer} {\tiny \dubbelop} {\tiny \dubbelop} \\
\textit{perfect} & 0.730 {\tiny (0.002)} {\tiny \dubbelneer} {\small $\circ$} {\small $\circ$} & 0.426 {\tiny (0.004)} {\tiny \dubbelneer} {\tiny \dubbelop} {\tiny \dubbelop} & 0.554 {\tiny (0.002)} {\tiny \dubbelneer} {\tiny \dubbelneer} {\tiny \dubbelneer} \\
\textit{cascading} & 0.696 {\tiny (0.008)} {\tiny \dubbelneer} {\tiny \dubbelneer} {\tiny \dubbelneer} & 0.320 {\tiny (0.006)} {\tiny \dubbelneer} {\tiny \dubbelneer} {\tiny \dubbelneer} & 0.415 {\tiny (0.014)} {\tiny \dubbelneer} {\tiny \dubbelneer} {\tiny \dubbelneer} \\
\textit{non-cascading} & 0.692 {\tiny (0.010)} {\tiny \dubbelneer} {\tiny \dubbelneer} {\tiny \dubbelneer} & 0.320 {\tiny (0.014)} {\tiny \dubbelneer} {\tiny \dubbelneer} {\tiny \dubbelneer} & 0.422 {\tiny (0.014)} {\tiny \dubbelneer} {\tiny \dubbelneer} {\tiny \dubbelneer} \\
\midrule
& \multicolumn{3}{c}{\textit{\acl{PDGD}}} \\
\midrule
\textit{perfect} & 0.752 {\tiny (0.001)} {\tiny \dubbelop} {\tiny \dubbelop} {\tiny \dubbelop} {\tiny \dubbelop} & 0.442 {\tiny (0.003)} {\tiny \dubbelop} {\tiny \dubbelop} {\tiny \dubbelop} {\tiny \dubbelop} & 0.592 {\tiny (0.000)} {\tiny \dubbelop} {\tiny \dubbelop} {\tiny \dubbelop} {\tiny \dubbelop} \\
\textit{cascading} & 0.730 {\tiny (0.003)} {\tiny \dubbelneer} {\small $\circ$} {\tiny \dubbelop} {\tiny \dubbelop} & 0.420 {\tiny (0.007)} {\tiny \dubbelneer} {\tiny \dubbelneer} {\tiny \dubbelop} {\tiny \dubbelop} & 0.563 {\tiny (0.003)} {\tiny \dubbelneer} {\tiny \dubbelop} {\tiny \dubbelop} {\tiny \dubbelop} \\
\textit{non-cascading} & 0.729 {\tiny (0.003)} {\tiny \dubbelneer} {\small $\circ$} {\tiny \dubbelop} {\tiny \dubbelop} & 0.424 {\tiny (0.005)} {\tiny \dubbelneer} {\tiny \dubbelneer} {\tiny \dubbelop} {\tiny \dubbelop} & 0.570 {\tiny (0.003)} {\tiny \dubbelneer} {\tiny \dubbelop} {\tiny \dubbelop} {\tiny \dubbelop} \\
\bottomrule
\end{tabular*}

\label{tab:main}
\vspace{-0.5\baselineskip}
\end{table*}

\section{Conclusion}
\label{sec:conclusion}

In this study, we have reproduced and generalized findings about the relative performance of \acf{DBGD} and \acf{PDGD}.
Our results show that the performance of \ac{PDGD} is reproducible under non-cascading user behavior.
Furthermore, \ac{PDGD} outperforms \ac{DBGD} in both \emph{ideal} and extremely \emph{difficult} circumstances with high levels of noise and bias.
Moreover, the performance of \ac{PDGD} in extremely \emph{difficult} circumstances is comparable to that of \ac{DBGD} in \emph{ideal} circumstances.
Additionally, we have shown that the regret bounds of \ac{DBGD} are not applicable to the ranking problem in \ac{LTR}.
In summary, our results strongly confirm the previous finding that \ac{PDGD} consistently outperforms \ac{DBGD}, and generalizes this conclusion to circumstances with extreme levels of noise and bias.

Consequently, there appears to be no advantage to using \ac{DBGD} over \ac{PDGD} in either theoretical or empirical terms.
In addition, a decade of \ac{OLTR} work has attempted to extend \ac{DBGD} in numerous ways without leading to any measurable long-term improvements.
Together, this suggests that the general approach of \ac{DBGD} based methods, i.e., sampling models and comparing with online evaluation, is not an optimally effective way of optimizing ranking models.
Although the \ac{PDGD} method considerably outperforms the \ac{DBGD} approach, we currently do not have a theoretical explanation for this difference.
Thus it seems plausible that a more effective \ac{OLTR} method could be derived, if the theory behind the effectiveness of \ac{OLTR} methods is better understood.
Due to this potential and the current lack of regret bounds applicable to \ac{OLTR}, we argue that a theoretical analysis of \ac{OLTR} could make a very valuable future contribution to the field.

Finally, we consider the limitations of the comparison in this study.
As is standard in \ac{OLTR} our results are based on simulated user behavior. 
These simulations provide valuable insights: they enable direct control over biases and noise, and evaluation can be performed at each time step.
In this paper, the generalizability of this setup was pushed the furthest by varying the conditions to the extremely difficult.
It appears unlikely that more reliable conclusions can be reached from simulated behavior.
Thus we argue that the most valuable future comparisons would be in experimental settings with real users.
Furthermore, with the performance improvements of \ac{PDGD} the time seems right for evaluating the effectiveness of \ac{OLTR} in real-world applications.

\vspace{-0.4\baselineskip}

\subsubsection*{Acknowledgements.}
This research was supported by 
Ahold Delhaize, 
the Association of Universities in the Netherlands (VSNU), 
the Innovation Center for Artificial Intelligence (ICAI),
and the Netherlands Organization for Scientific Research (NWO)
under pro\-ject nr
612.\-001.\-551.
All content represents the opinion of the authors, which is not necessarily shared or endorsed by their respective employers and/or sponsors.

 \bibliographystyle{splncs04}
 \bibliography{ecir2019-oltr-optimization}
 
 \end{document}